\documentclass[letterpaper, 10 pt, conference]{ieeeconf}  
\IEEEoverridecommandlockouts         
\usepackage{times} 
\usepackage{url}
\usepackage{float}
\usepackage{amsmath} 
\usepackage{algpseudocode}
\usepackage{subfig}
\usepackage{graphicx}
\usepackage{amssymb}  
\title{\LARGE \bf Adversarial Path Planning for Optimal Camera Positioning}
\author{Gaia Carenini and Alexandre Duplessis$^{*}$
\thanks{$^{*}$Gaia Carenini and Alexandre Duplessis are Master Students of the Department of Computer Science at ENS-PSL Research University, Paris, France. They contributed equally to the realization of this work. Their emails are:
{\tt\small name.surname@ens.psl.eu}}%
}
\begin{document}
\maketitle
\thispagestyle{empty}
\pagestyle{empty}

\begin{abstract}
 \noindent The use of visual sensors is  flourishing, driven among others by the several applications in detection and prevention of crimes or dangerous events. While the problem of optimal camera placement for total coverage has been solved for a decade or so, that of the arrangement of cameras maximizing the recognition of objects "in-transit" is still open. The objective of this paper is to attack this problem by providing an adversarial method of proven optimality based on the resolution of Hamilton-Jacobi equations. The problem is attacked by  first assuming the perspective of an adversary, i.e. computing explicitly the path minimizing the probability of detection and the quality of reconstruction. Building on this result,  we introduce an optimality measure for camera configurations and perform a simulated annealing algorithm to find the optimal camera placement.
\end{abstract}
\section{Introduction}
\noindent Networks of cameras, or generally of visual sensors, are widely used throughout industrial processes, detection and prevention
of crimes or dangerous events, military purposes and so forth. The vast availability of different types of cameras, the decreasing cost of associated hardware, together with an increasing need for such systems are among the reasons intriguing more and more researchers to focus in this field.\\
A crucial concern raised by the design of a camera network is positioning optimally individual visual sensors under a set of default constraints. In fact, resulting visual measurements can be made significantly more accurate by selecting a suitable configuration established through a proper mathematical
model. It's clear that different visual tasks have fairly distinct requirements, e.g., a multi-view reconstruction task stands in need of a minimum number of video sensors with predetermined ranges of angular separation, instead, some aggregate video sensor network must be fault-tolerant to camera drop out and still layouts of sensors in video
sensor networks should assure a minimum level of image quality in order to get a sufficient  resolution, depth of field, etc.
Independently from the specific task, resolution is always a fundamental and primary information bottlenecks for vision applications.\\ 
The problem of automating the camera network design process for attaining highly accurate measurements has received comparatively little attention given its practical importance.\\
Our goal is to address the problem of camera placement to optimize the aggregate observability. One possible application of this research is the
development of a design tool for surveillance camera
placement in areas of high trafﬁc, where each subject
may take a different path through the area. This work
assumes the cameras are statically mounted to view an
area. Optimizing the observability of such a system means jointly maximizing the power of observation of the cameras for the area of interest.\vspace{0.2cm}\\
\textsc{Related work}$\xrightarrow{}$ Suitable camera placement for the purpose of optimizing the sensors
ability to capture information about a desired environment or task has
been studied extensively. In \cite{BookArt}, O’Rourke provides an in-depth
theoretical analysis of the problem of maximizing camera coverage
of an area, where the camera fields of view do not overlap (the so-called “art gallery” problem). Several other results have been published in this direction, e.g., \cite{ArtGal,ArtGallery}.\\
In recent years, research has sought to extend the framework
to include limited field of view cameras and to incorporate resolution
metrics into the formulation (see \cite{cov1,cov, cov2}). More specifically, in \cite{relWork},  the art gallery framework is refined by introducing a resolution quality metric. Moreover in \cite{relwork1}, the formulation of the minimum guard coverage art gallery problem was extended in order to incorporate minimum-set cover. In the same work,
reduced upper bounds were derived for two cases of exterior visibility for two- and three-dimensions.\vspace{0.2cm}\\
\textsc{Contribution}$\xrightarrow{}$ Our method differs from the art gallery framework in various salient aspects. First, we study the best attacks to visibility conditioned to a given camera configuration and destination, i.e., we find out the paths minimizing the observability of the object "in-transit" from a couple of selected positions. 
This is done by modelling this task as an optimal motion planning problem that can be solved by applying the technique presented in \cite{method}. 
We then derive an optimality measure that can be used to assess a configuration. We finally use a simulated annealing based algorithm to find a near-optimal camera placement according to this measure.
\section{Preliminaries}
\noindent We start by reducing the design of "attacks to observability" to a problem of optimal motion planning in a space presenting an anisotropic field of velocities; there the goal is to reach a final
position $(x_f, y_f)$ from a start position $(x_0, y_0)$ in a minimum time, while avoiding obstacles and minimizing the risk of being recognized.\\
For sake of simplicity, we assume that all the cameras are punctiform and we consider a finite 2D environment defined as $\mathcal{R}\subset[a,b]\times[c,d]$ with $a,b,c,d \in \mathbb{R}$. We define an obstacle, $obs$, as a proper subset of $\mathcal{R}$ and we consider the set of the obstacles $\mathcal{O}$. We define each camera as a tuple $C_i=(p_i, \alpha_i, r_i)$ in which $p_i$ is of the form $(a_i,b_i,\beta_i)$ where $(a_i,b_i)$ are the coordinates of the camera $C_i$, $\beta_i$ is the angle that the first ray of the vision field of $C_i$ defines with the vertical according to the standard reference system, $\alpha_i$ is the angular opening of the camera, and $r_i$ is a function describing the resolution that the camera has of an object when the distance from it changes. The vision field is the space swept by any vector moving from the position $\beta_i$ to the position $\beta_i+\alpha_i$, we call it $\mathcal{F}_i$.  For the model proposed below, we will assume that the recognition is jointly inversely proportional to the distance from the camera and directly proportional to the time spent in the field of view of the camera. Other assumptions could be added in a straightforward manner in the model presented that is kept simple for sake of clarity.\\ Fix the start at $(x_0,y_0)\in \mathcal{R}$, the destination point in $(x_f,y_f)\in \mathcal{R}$ and $N$ cameras $\mathcal{C}=\{C_i\}_{i=1}^N$, 
we model cameras visual field as an anisotropic speed field $\overrightarrow{W}$ that depends on $(x_f,y_f)$ as follows:
$\forall (x,y)\in \mathcal{R}\hspace{0.3cm} \overrightarrow{w}(x,y)$ is either 0 if $(x,y)\notin \bigcup_{i\in [1,N]} \mathcal{F}_i$ or as the vector that has direction the opposite of the conjunction of $(x,y)$ and $(x_f,y_f)$ and has module $1/dist ((a_i,b_i),(x,y))^2$.\\
Under this assumption, the  speed of the motion that we consider is defined by the equations:
\begin{equation}\label{speedField}
     \begin{cases}
      \dot{x}(t)=(V_c+w(x,y))\sin(\theta(t))\\
      \dot{y}(t)=(V_c+w(x,y))\cos(\theta(t))
    \end{cases}
\end{equation}
where (x,y) is the mobile position, $\theta$ is the heading angle relative to north direction and $V_c$ is a constant finite speed of the object "in transit" that has direction corresponding to the conjunctive from the start position and the destination and constant module $c$.\\
With the formalism introduced above, we can define 
the resolving optimization problem that can be written as:
\begin{equation}\label{pr}
     \begin{cases}
      \min\hspace{0.4cm}(t_f-t_0)\\
      s.t.\hspace{0.5cm}\dot{x}(t)=(V_c +w(x,y))\sin(\theta(t))\\
      \hspace{1 cm}\dot{y}(t)=(V_c+w(x,y))\cos(\theta(t))\\
      \hspace{1cm}(x(t_0), y(t_0))=(x_0,y_0)\\
      \hspace{1cm}(x(t_f), y(t_f))=(x_f,y_f)\\
    \end{cases}
\end{equation}
We observe that the control parameter of (\ref{pr}) is the the heading angle $\theta$ and that the solution of the optimization problem is therefore finding the $\theta$ that over time  minimize the total travel time.\\
For sake of simplicity, we consider as control variable the unit vector naturally associated with it, i.e. $\overrightarrow{a}(t)=(sin(\theta(t)), cos(\theta(t)))$. In this case, the problem can be restated simply as: 
\begin{equation}\label{prob2}
    \begin{cases}
     \min_{\overrightarrow{a}}\hspace{0.4cm}(t_f-t_0)\\
     s.t\hspace{0.4cm}\dot{X}=f(X(t),\overrightarrow{a}(t))\\
      \hspace{0.8cm}X(t_0)=X_0\wedge X(t_f)=X_f
    \end{cases}
\end{equation}
where $X$ is the  position of the object "in-transit" and $f(X(t),\overrightarrow{a}(t))$ is the real speed of the mobile at time $t$.\\
 The optimal control problem introduced (\ref{prob2}) is classical and we can observe how the corresponding Hamilton-Jacobi equation is given by:
\begin{equation}\label{eq}
    \max_{\overrightarrow{a}\in A}\{-\nabla u(X), f(X,\overrightarrow{a})\}=1
\end{equation}
where $u(X)$ represents the minimum time to reach the destination starting from the point $X$.
\section{Motion planning}\label{path_planning}
\noindent In this section, we discuss the resolution of (\ref{eq}). The same issue has been solved in \cite{method} in the context of optimal motion planning in presence of wind. For seek of completeness, we outline below the method.\\
The idea behind the resolution is to decompose
recursively the problem into linked sub-problems as it happens in dynamic programming. More specifically, the resolution can be seen as a front expansion problem, where the wavefront
represents the minimum time to reach the arrival point. The computation is based on the classical Huygen’s principle, which that
states "every point reached by a wavefront becomes a source
of a spherical wavefront". The evolution of the wavefront is given by:
\begin{equation}
    ||\nabla u(X)||F \left( X, \frac{\nabla u(x)}{||\nabla u(X)||} \right)=1
\end{equation}
where $F(X, \overrightarrow{n})$ is the front speed in the direction $\overrightarrow{n}$  of the outward unit vector normal to the front at point $X$.\\
Through some algebraic manipulation (see \cite{method}), the optimal path problem can therefore be written as a front
expansion problem where the speed of the wavefront is given by:
\begin{equation}
    F(X,\overrightarrow{n})= \max_{\overrightarrow{a}}\{-\overrightarrow{n}.f(X,\overrightarrow{a})\}
\end{equation}
To design the optimal path between the departure point and
the final point, it is sufficient to exploit the characteristics of the
Hamilton-Jacobi PDE.\\
Several methods exist for finding an approximation of the solution of these Hamilton-Jacobi equations, in this context, we apply the so-called \emph{ordered upwind algorithm}.
Presented in \cite{Upwind}, this technique was proven to converge to a weak solution of the PDE, in particular to the \emph{viscosity} one.
Its basic principle is to avoid useless iterations, common in Dijkstra-like methods, thanks to a careful use of the information about the characteristic directions of the PDE.\\
The first step consists in computing the value function, $u$, considering a $2 \mathrm{D}$ nonregular triangular mesh. The method applied is described extensively in \cite{method}. Other methods exist, e.g. semi-Lagrangian and Eulerian discretization \cite{Upwind}, however both of them have disadvantages requiring 
multiple local minimization and finding the roots of a non-linear equation respectively.\\
In our case, the speed of the wavefront $F$ has a closed form, and the value can be computed using a finite-differences upwind formula of the Hamilton-Jacobi equation.\\ 
By modelling the object "in-transit" speed as $f(\mathrm{X}, a)=V_{a} a+W$, the speed of the wavefront is equal to:
\begin{equation}
   F(\mathrm{X}, n)=V_{a}-\langle n, W\rangle 
\end{equation}
For more details, read \cite{com}. Fixed this definition of the wavefront speed, to the problem is applied an upwind finite-difference discretization on the simplex $\left(\mathrm{X}, \mathrm{X}_{j}, \mathrm{X}_{k}\right)$.
The associated Hamilton-Jacobi equation becomes:
\begin{equation}
\left\|P^{-1} w(\mathrm{X})\right\|^{2} V_{a}^{2}=\left(1+\left\langle P^{-1} w(\mathrm{X}), W\right\rangle\right)^{2} 
\end{equation}
where the vector $P^{-1} w(\mathrm{X})$ is the discretization of $\nabla u(\mathrm{X})$ from the directional derivatives of $u$ in the directions defined by the edges of the simplex $\left(\mathbf{x}, \mathbf{x}_{j}, \mathbf{x}_{k}\right)$. This equation has the property of being  quadratic and has the following form:
\begin{equation}
A v_{\mathbf{x}_{j} \mathbf{x}_{k}}^{2}(\mathrm{X})+B v_{\mathbf{x}_{j} \mathbf{x}_{k}}(\mathrm{X})+C=0 
\end{equation}
where the coefficients are given by:
\begin{small}
$$
\begin{aligned}
& A=V_{a}^{2}\left\langle P^{-1} \alpha, P^{-1} \alpha\right\rangle-\left\langle P^{-1} \alpha, W\right\rangle^{2} \\
& B=2 V_{a}^{2}\left\langle P^{-1} \alpha, P^{-1} \beta\right\rangle-2\left\langle P^{-1} \alpha, W\right\rangle\left(\left\langle P^{-1} \beta, W\right\rangle+1\right) \\
& C=V_{a}^{2}\left\langle P^{-1} \beta, P^{-1} \beta\right\rangle-\left[\left\langle P^{-1} \beta, W\right\rangle+1\right]^{2}
\end{aligned}
$$
\end{small}

\noindent The value $v_{\mathbf{x}_{j} \mathbf{x}_{k}}$ is then computed by the classical resolution formula of quadratic equation. To ensure that $v_{\mathbf{x}_{j} \mathbf{x}_{k}}$ is a good approximation of the value function, $u$, at the point $\mathrm{X}$, the characteristic direction for the mesh point $\mathrm{X}$ needs to lie inside the simplex $\left(\mathrm{X}, \mathrm{X}_{j}, \mathrm{X}_{k}\right)$.
The optimal trajectory is built by moving from the initial point to the destination point along the characteristic direction determined by:
\begin{equation}
\frac{d \mathrm{X}}{d t}=-V_{a} \frac{\nabla u(\mathrm{X})}{\|\nabla u(\mathrm{X})\|}+W(\mathrm{X})  
\end{equation}
The computational complexity of this algorithm is $O(\gamma N \log N)$, where $N$ is the number of mesh points and $\gamma$ the anisotropy ratio (see \cite{Upwind}).\vspace{0.2cm}\\
\textsc{Obstacle avoidance}$\rightarrow$ Two approaches can be used. The first one, derived from \cite{method}, consists in decreasing the speed of propagation of the wavefront in the parts of the environment corresponding to obstacles. In this way, we register a growth of the value function $u$ that penalizes the passage through these areas. We define a map of values $\xi$
as a function of the obstacles, $\mathcal{O}$. The values $\xi$ are between $0$ and an upper bound $\xi_{max}$. The scaled values $\xi$ are then exploited in order to slow down the wavefront speed as follows: $(1-\epsilon)F(X,n)$.
The maximum value $\xi_{max}$ needs to be less than 1 to keep the
wavefront speed strictly positive.\vspace{0.2cm}\\
Another approach, which we find simpler - especially from a computational viewpoint, is based on the fact that the Hamilton-Jacobi equation's resolution method proposed above can be seen as a shortest-path algorithm in the graph whose vertices are the mesh points, and edges represent the neighborhood relationship. Thus taking in account obstacles can done by simply deleting the obstacle's vertices from the graph.
\\

\textsc{DL approach}$\xrightarrow{}$ The formalization of the problem proposed leads to the resolution of a Hamilton-Jacobi equation. Problems involving these kind of equations have been studied in several areas of mathematics, numerical computing and more recently in deep learning and several networks have been designed including \cite{Deep1, deep2}. In this work, we do not pursue this direction, having decided to privilege convergence guarantees rather then speed. 
\section{Optimal camera placement}
\noindent In this section, fixed a camera configuration $C=\{C_i\}_{i=1}^N$ , we provide a measure for assessing how much effective this configuration is when it comes to preventing an object "in-transit" to go unnoticed. For sake of simplicity, we start assuming that the adversary has fixed starting position $(x_0, y_0)$ and final destination $(x_f, y_f)$. A fairly immediate generalization consists in averaging the optimality measure, defined below, on every pair of positions $((x_0, y_0), (x_f, y_f))$ that an adversary could take (or more realistically on a sampling of them).
We define our objective so that it takes into account the integral of the "portion" of the path, returned by applying the ordered upwind algorithm (from now on, called algorithm $\mathcal{A}$), that intersects at least one of the camera's field of view. A detail to notice is that we normalize this quantity over the distance between the initial position and the end position, since this distance should not influence the complexity of an adversarial path. Therefore, for any parametrization $\gamma:[0,1]\rightarrow \mathcal{R}$ of a valid path (i.e. a path avoiding obstacles), we define:
\begin{equation}\label{meas}
    \mathcal{L}_{(p_i)_{1\leq i \leq N}}(\gamma) = \int_0^1 1 + \eta \mathbf{1}\left(\gamma(t)\in \underset{1\leq i \leq N}{\bigcup} \text{scope}(C_i)\right) \text{d} t
\end{equation}
where $\eta$ controls the tradeoff between path length and camera visibility, and scope$(C_n)\subset \mathcal{R}$ is the set of points of $R$ that a camera $C_n$ has in its field of view ; more formally, it is defined as:
\begin{equation}
\begin{split}
    \text{scope}(C_i) &= \Bigg\{ (x,y)\in \mathcal{R} \text{ s.t. } \\ &\beta_i \leq \arctan \left(\frac{y-b_i}{x-a_i}\right) \leq \beta_i + \alpha_i  \\ 
    & \text{and } ](a_i,b_i), (x,y)[ \cap \mathcal{O} = \emptyset \Bigg\}
\end{split}
\end{equation}
where we have used the natural extension of $\arctan$ to  $\overline{\mathbb{R}}$.\vspace{0.2cm}\\
\textsc{Simulated Annealing}$\xrightarrow{}$ Given  (\ref{meas}), the initial problem can be restated as follows: 
\begin{equation}\label{problem}
    \underset{(x_i,y_i,\theta_i)_{1\leq i \leq N}}{\min} \mathcal{L}(\mathcal{A}(\{(x_i, y_i, \theta_i)_i\}))
\end{equation}
for $N$ a given number of cameras.

An important remark is that we actually do not need to compute $ \mathcal{L}(\mathcal{A}(\{(x_i, y_i, \theta_i)_i\})$ since a similar measure is already computed during the path computation, i.e. $u(x_f)$. Therefore in practice the simulated annealing algorithm is performed using $u(x_f)$ as optimality measure.

The optimization required by (\ref{problem}) cannot be performed using classical methods, e.g. gradient descent, both because of the complexity of gradients' estimation, an because of the existence of several local minima (see \cite{resolution}).
For this reason, we decided to use simulated annealing (SA, see \cite{Potential}), an effective method for approximating global optima. The principle behind this technique was inspired by annealing in metallurgy and consists in proposing a new potentially optimizing candidate at each step and accepting it always if it scores better, but still with some probability $p$ otherwise.
In our framework, at each step, SA randomly selects one of the $n$ cameras and sets its parameters randomly. The new configuration obtained $C^{t+1}$ is then evaluated thanks to our optimality measure. If the new configuration scores better than the previous one $C^{t}$, we accept it. Otherwise we still accept the new configuration with some probability. This is equivalent to  accepting the new configuration with probability:
 \begin{equation}
 \begin{split}
    p(\text{accept} &| C^{t}, C^{t+1}, T) \\
    &= \min \{ 1, \exp (\frac{\mathcal{L}(\mathcal{A}(C^{t+1})) - \mathcal{L}(\mathcal{A}(C^{t}))}{T})\}
    \end{split}
\end{equation}

where $T$ is called the annealing temperature, and is typically chosen high at the beginning and then decreased gradually. In this work, we adopt a linear schedule for $T$. The complete pseudocode for the algorithm is the following.\vspace{0.2cm}
\begin{algorithmic}[1]
\Procedure{SA}{$T_0$}
\State $C \gets \text{RANDOM}()$
\For{$k=0$ to $K$}
    \State $T \gets \max (0, T_0 (1-\frac{k+1}{K}))$
    \State $C_\text{prop} \gets \text{RANDOM}()$
    \If{$p(\text{accept} | C, C_{prop}, T) >$ $\text{random}(0,1)$}
        \State $C \gets C_{\text{prop}}$
    \EndIf
\EndFor
\State \textbf{return} $C$
\EndProcedure
\end{algorithmic}
\vspace{0.2cm}
Concerning optimality, SA is guaranteed to converge to the global optimum in a finite time if the candidates and the temperature satisfy some well-known weak conditions \cite{simulatedproof}. From a practical point of view, since the search-space is infinite, convergence to the global minimum may be slow, but results are still very satisfying and may be further improved by using local optimization methods.

\section{Results}
\begin{figure}[H]
    \centering
    \includegraphics[scale=0.50]{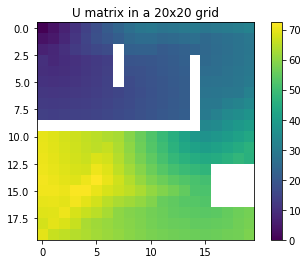}
    \caption{$U$ matrix in $20\times 20$ grid}
    \label{u_special}
\end{figure}

\begin{figure}[H]
    \centering
    \includegraphics[scale=0.50]{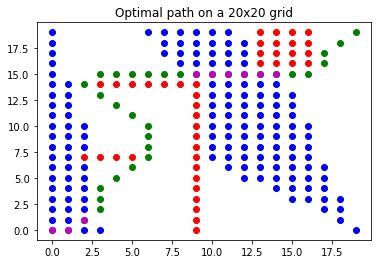}
    \caption{Visualization of optimal path in a $20\times 20$ grid (Obstacles points are plotted in red, camera visible points in blue, path points in green, and path points visible by at least one camera in purple.)}
    \label{p_special}
\end{figure}

\begin{figure*}
\begin{tabular}{ccc}
  \includegraphics[width=.3 \textwidth]{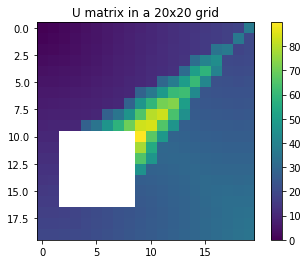} &   \includegraphics[width=.3 \textwidth]{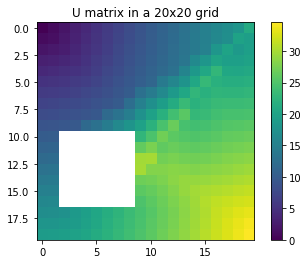} &
  \includegraphics[width=.3 \textwidth]{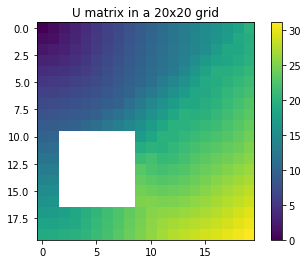}\\[6pt]
 \includegraphics[width=.3 \textwidth]{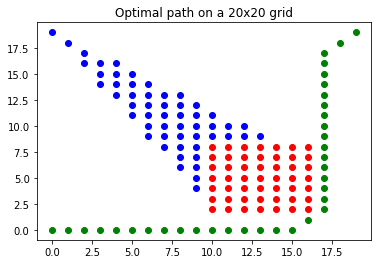} &   \includegraphics[width=.3 \textwidth]{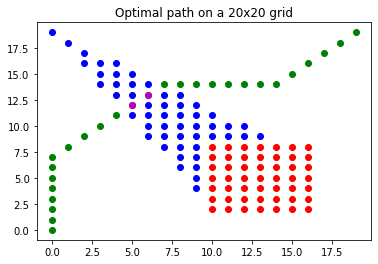} &
  \includegraphics[width=.3 \textwidth]{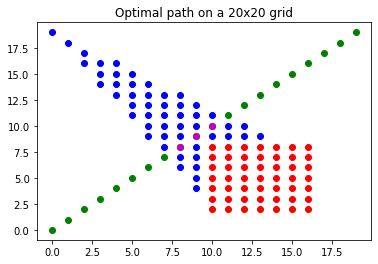}\\[6pt]
\end{tabular}
\caption{$U$ matrices and optimal path visualization with increasing weight put on the path length compared to camera visibility from left to right (same colors as previous figures)}
\label{tradeoff}
\end{figure*}

\noindent After a sequence of tests, the model developed seems to achieve fairly good performance. We discuss them below making a distinction among the results concerning path planning and camera placement.\vspace{0.2cm}\\
\textsc{Path planning}$\rightarrow$ We have implemented the path planning algorithm fully described in \ref{path_planning}. For seek of clarity, we show the results in a specific case, where the norm of the vector field $w$ does not depend on the distance with respect to camera position ; this is motivated by the will of visualizing easily the trade-off among minimizing the path length and avoiding on cameras.\\
\noindent Figures \ref{u_special} and \ref{p_special} show an example of output of our algorithm in a complex environment.
\noindent Figure \ref{tradeoff} shows how the tradeoff between path length and camera visibility influences the optimal path. The results can be interpreted very intuitively\footnote{Because we allow exclusively four moves from each position in the grid, straight lines in the continuous space are either parallel to an axis or a bisector of them that is achieved thanks to two straight lines, giving a longer path. This is a minor issue that does not affect the placement algorithm for a fixed number of cameras. However a potential improvement to the current implementation could consist in "post-replacing" the paths by trying to directly join points not separated by a camera scope (if they do not intersect an obstacle).}, since the more we privilege camera avoidance (compared to path length minimization), the more the agent tries to spend less time in the camera scope, and thus either makes a detour around the obstacle or (in the intermediate case) goes closer to the camera position \footnote{Due to the discretization, the number of points in the scope of the camera is not an increasing function of the distance to the camera. This is a minor problem that affects neither the general behavior of the algorithm, nor the following camera placement, especially when using high-dimension grids.}.

\vspace{0.2cm}
\textsc{Camera placement}$\rightarrow$ The results of simulated-annealing algorithm for camera placement seem convincing as well. In fact, despite slow computation, the results obtained are easily interpretable.

\begin{figure}[H]
    \centering
    \includegraphics[scale=0.50]{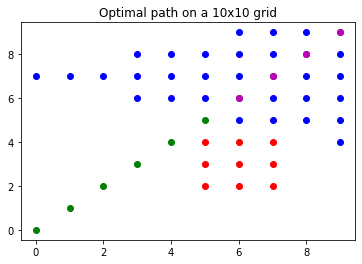}
    \caption{Optimal placement of $1$ camera and associated path}
    \label{optimal_p}
\end{figure}

\noindent Figure \ref{optimal_p} shows the optimal placement of one camera to prevent an agent going from bottom left corner to top right corner from avoiding the camera, with one obstacle. Positioning the camera as found is in fact intuitively optimal.

\section{Conclusion}
\noindent This work proposes a new multi-camera placement modeling set-up to support the design of network of visual sensor. The model allows for taking into account important constraints that are involved in computer-vision applications operated on the cameras' recordings. In fact, the anisotropic speed field is easily adaptable by changing the definition of $w$ (in equation \ref{speedField}), so that it explicitly accounts for resolution or any other desired property. The formalism reduces the camera placement problem to a coupled optimization that builds partially on prior work in the field of motion planning. Furthermore, in addition to an optimal camera placement algorithm, we provide a way to assess the optimality of a given camera network.\vspace{0.2cm}\\ 
\textsc{Limitations and Further Work}$\xrightarrow{}$
This work presents a few limitations that we list below (in decreasing order of importance) to be addressed in future work. First of all, we have not taken into account the triangulation constraint. This could be easily solved by adapting the SA algorithm. A second improvement that could be made would be to allow a non-fixed number of cameras during the optimal placement process. This could be solved by introducing a joint optimization of the number of cameras and their parameters through transdimentional SA \cite{transdimentionalsa}. One last limitation of our work is related to the speed of computation that is quite modest, and the algorithm could be considerably accelerated (for instance by discretizing the positions of the cameras with a step size chosen according to the size of the mesh and a "characteristic size" of the obstacles). However given the type of applications, a high computing speed is not necessary since the computation must be done only once for a given enviornment.

\end{document}